\documentclass[]{spie}

\usepackage{graphicx,capt-of,wrapfig}
\usepackage{amsmath,amssymb,hyperref}

\newcommand\farcs{\mbox{$.\!\!^{\prime\prime}$}}%
\newcommand\arcs{\mbox{$^{\prime\prime}$}}%

\title{MagAO: Status and Science} 

\author{
	Katie M. Morzinski\supit{a*},
	Laird M. Close\supit{a},
	Jared R. Males\supit{a$\dagger$},
	Phil M. Hinz\supit{a},
	Simone Esposito\supit{b},
	Armando Riccardi\supit{b},
	Runa Briguglio\supit{b},
	Katherine B. Follette\supit{c},
	Enrico Pinna\supit{b},
	Alfio Puglisi\supit{b},
	Jennifer Vezilj\supit{a},
	Marco Xompero\supit{b},
	and
	Ya-Lin Wu\supit{a}
	\skiplinehalf
	\supit{a} Steward Observatory, University of Arizona, Tucson AZ 85721, USA; \\
	\supit{b} INAF-Osservatorio Astrofisico di Arcetri, Largo E. Fermi 5, I-50125 Firenze, Italy; \\
	\supit{c} Kavli Institute for Particle Astrophysics \& Cosmology, Physics Department, Stanford University, Stanford, CA, 94305
}

\authorinfo{\supit{*}Contact: ktmorz@arizona.edu \\ \supit{$\dagger$}NASA Sagan Fellow}
 
\begin{document} 
\maketitle 

\begin{abstract}
``MagAO'' is the adaptive optics instrument at the Magellan Clay telescope at Las Campanas Observatory, Chile.  MagAO has a 585-actuator adaptive secondary mirror and 1000-Hz pyramid wavefront sensor, operating on natural guide stars from $R$-magnitudes of -1 to 15.  MagAO has been in on-sky operation for 166 nights since installation in 2012.  MagAO's unique capabilities are simultaneous imaging in the visible and infrared with VisAO and Clio, excellent performance at an excellent site, and a lean operations model.  Science results from MagAO include the first ground-based CCD image of an exoplanet, demonstration of the first accreting protoplanets, discovery of a new wide-orbit exoplanet, and the first empirical bolometric luminosity of an exoplanet.  We describe the status, report the AO performance, and summarize the science results.  New developments reported here include color corrections on red guide stars for the wavefront sensor; a new field stop stage to facilitate VisAO imaging of extended sources; and eyepiece observing at the visible-light diffraction limit of a 6.5-m telescope.
We also discuss a recent hose failure that led to a glycol coolant leak, and the recovery of the adaptive secondary mirror (ASM) after this recent (Feb.\ 2016) incident.
\end{abstract}

\section{INTRODUCTION}
The Magellan Adaptive Optics (``MagAO'') instrument has been in operation at the 6.5-m Clay Telescope at Las Campanas Observatory, Chile since Fall 2012.  MagAO uses an adaptive secondary mirror (ASM) for wavefront correction as proposed by Salinari\cite{1994ESOC...48..247S}, and is part of the second generation to do so (the first-generation ASM on-sky being at the MMT on Mt.\ Hopkins, Arizona\cite{2010SPIE.7736E..2BB}).  The wavefront sensor (WFS) is a modulating pyramid WFS, which was proposed by Ragazonni\cite{1996JMOp...43..289R} and demonstrated regularly on-sky at the LBT on Mt.\ Graham, Arizona\cite{2010SPIE.7736E..09E,2010ApOpt..49G.174E}.  MagAO uses natural guide stars (NGS) to correct single-conjugate fields, on-axis or up to $\sim$100$\arcs$ off-axis.

Magellan AO has two science cameras: the visible-light imager VisAO (600--1000 nm)\cite{2013PhDT.......261M,2014ApJ...786...32M} and the infrared imager Clio (1--5 $\mu$m)\cite{2003SPIE.4839.1154H,2004SPIE.5492.1561F,2006SPIE.6269E..0US,2010ApJ...716..417H,2015ApJ...815..108M}.  The two science cameras receive their respective light simultaneously via beam splitters, enabling observations in the visible and infrared (IR) wavelengths at the same time.  That is, all of the infrared light is sent to Clio at all times, whereas a selectable proportion of the visible light is split between VisAO and the visible-light WFS.

The instrument is operated in a block schedule where it is mounted on the telescope typically twice per year for 4--6 weeks at a time.  Visiting astronomers come to take their own data, while the MagAO team runs AO and collaborates to optimize the science.  The instrument is described in more detail in previous works\cite{2012SPIE.8447E..0XC,2012ApJ...749..180C,2014SPIE.9148E..04M}.

\section{ADAPTIVE OPTICS}
\begin{wrapfigure}{r}{0.36\linewidth}
	\vspace{-18pt}
	\begin{center}
		\begin{tabular}{c}
			\includegraphics[width=\linewidth]{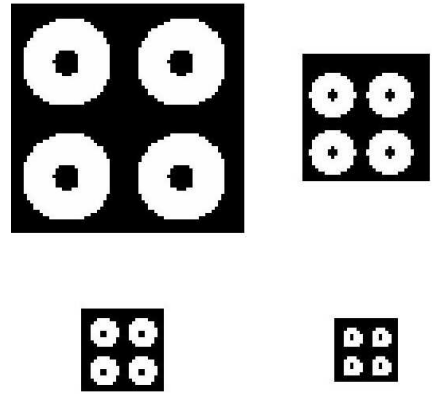}
		\end{tabular}
	\end{center}
	\vspace{-12pt}
	\caption[Pupils]
	{ \label{fig:pupils}
		Software registration of the pyramid pupils.
		The pupil illumination is aligned to $<$0.1 pixel via the camera lens loop.
		\textit{Top, left to right:} Bins 1 and 2.
		\textit{Bottom, left to right:} Bins 3 and 4.
	}
	\vspace{-18pt}
\end{wrapfigure}

MagAO's wavefront corrector is a 585-actuator adaptive secondary mirror, of which 540 actuators are illuminated.  The effective subaperture size is $d=24$ cm, taking into account the circular actuator arrangement on the ASM and the square subaperture arrangement on the WFS CCD\cite{2010SPIE.7736E..3HQ}.
The pyramid WFS has 27x27 subapertures appropriate for $R \lesssim$9, and can be readily modified down to 15x14, 9x9, or 6x6 subapertures via binning the CCD pixels (Fig.~\ref{fig:pupils}).
Corresponding modal control is via 21--378 Karhunen-Lo{\`e}ve modes, at frequencies from 105--1000 Hz.  Natural guide stars down to $R$$\sim$15 give good correction into the longer IR wavelengths, while NGS down to $R$$\sim$12 give good correction in the visible wavelengths.

During operation, a look-up table takes as input the NGS magnitude and a qualitative seeing estimate, and gives as output: the optimum configurations for the WFS frequency and binning; the corresponding amplitude and frequency of the modulation about the pyramid tip; the CCD gain; the reconstructor; and which pixels on the CCD should be illuminated by the pyramid pupils (Fig.~\ref{fig:pupils}).  The AO operator aligns the pupil illumination to the expected pupils and closes the loop with a low-order 10-mode loop to refine alignment\cite{2014SPIE.9148E..04M}.  An automated gain optimization script is then run that minimizes the wavefront error (WFE) for each gain for tip/tilt, low orders, and high orders.  Finally, the loop is closed fully to begin science observations.

\subsection{AO Performance}

\begin{figure}[htb]
\begin{minipage}{0.6\linewidth}
	\begin{tabular}{c}
	\includegraphics[height=0.9\linewidth,angle=90]{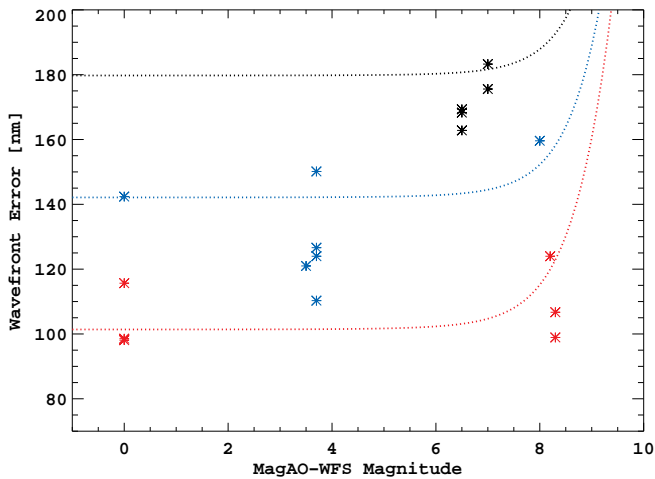}
	\end{tabular}
	\caption[Current WFE Performance]
	{ \label{fig:wfe}
		MagAO performance vs.\ NGS magnitude. The dotted lines are predictions from detailed analytic performance modeling.  The asterisks are on-sky measurements at various wavelengths.  The red curve/points correspond to good (1st quartile) conditions, blue corresponds to median conditions, and black to poor (3rd quartile).  On-sky points from the MagAO VisAO camera\cite{2016males}.
	}
\end{minipage}
\begin{minipage}{0.38\linewidth}
\captionof{table}{\label{tab:wfe}Turbulence model for Cerro Manqui (the Magellan telescopes' site).  Details given in Males \textit{et al.}, this proceedings\cite{2016males}.}
\vspace{10pt}
	\begin{tabular}{lrrrr}
		Quartile & Seeing$^*$ & $v_w$$^\star$ & $r_0$$^\dagger$ & $\tau_0$$^\ddagger$ \\
		 & FWHM & [m/s] & [cm] & [ms] \\
		\hline
		1st (red) & 0\farcs51 & 9.4 & 20.0 & 6.7 \\
		2nd (blue) & 0\farcs65 & 18.7 & 15.9 & 2.6 \\
		3rd (black) & 0\farcs81 & 23.4 & 12.7 & 1.7 \\
		\hline
		\multicolumn{5}{l}{$^*$At $\lambda$= 500 nm.} \\
		\multicolumn{5}{l}{$^\star$$C_n^2$-weighted wind speed.} \\
		\multicolumn{5}{l}{$^\dagger$$r_0 = \lambda / $FWHM.} \\
		\multicolumn{5}{l}{$^\ddagger$$\tau_0 = 0.31 r_0 / v_w$.} \\
		\multicolumn{5}{p{\linewidth}}{{\it N.B.}: $\tau_0$ calculation only valid for cases in which per-quartile seeing and per-quartile wind speed occur simultaneously.}
	\end{tabular}
\end{minipage}
\end{figure}

Figure~\ref{fig:wfe} shows MagAO performance as a function of NGS magnitude.
Table~\ref{tab:wfe} shows the associated atmospheric turbulence parameters used in generating the model performance curves\cite{2016males}, as well as a calculation of the corresponding temporal coherence length.
Optimum performance is achieved for stars brighter than $R$$\sim$8.
When the wind speed and seeing are in the first quartile, the temporal coherence length of $\sim$6.7 ms allows for excellent sustained AO performance.

\subsection{WFS Color Correction}

\begin{figure}[htb]
	\begin{center}
		\begin{tabular}{c}
			\includegraphics[width=0.49\linewidth]{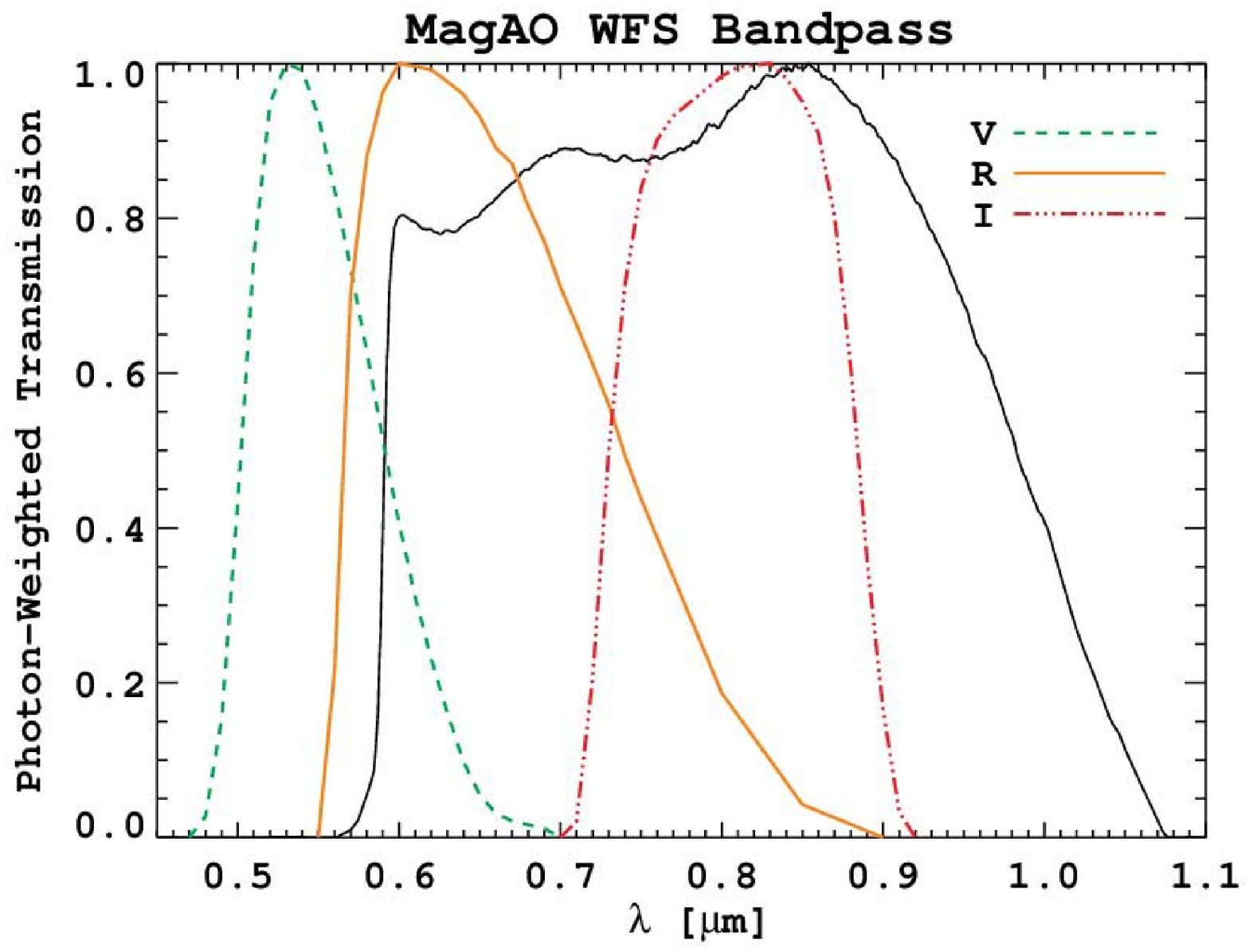}
			\includegraphics[width=0.49\linewidth]{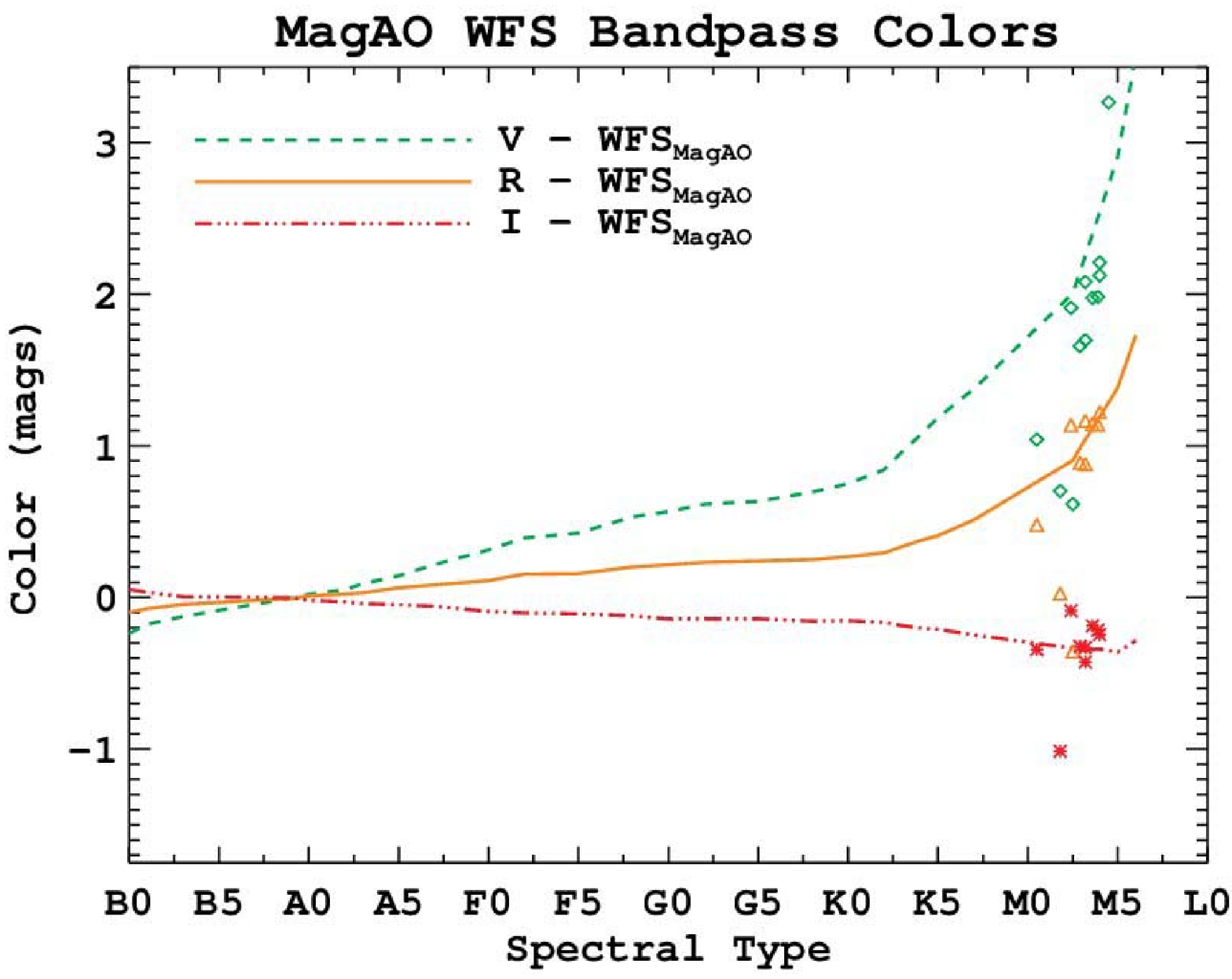}
		\end{tabular}
	\end{center}
	\caption[MagAO bandpass]
	{ \label{fig:bandpass}
		\textit{Left:} The bandpass of MagAO's WFS (black line) is as wide as and somewhat redder than $R+I$.
		\textit{Right:} Color corrections for stars of late spectral types are plotted against on-sky measurements of the magnitude by the WFS.
	}
\end{figure}

The look-up table that selects the AO configurations for each new guide star is based on the astronomer's estimate of the $R$- or $I$-band magnitude.
For faint and red targets, however, the estimate consistently gives poorly-optimized configurations.
This is because MagAO's pyramid WFS has a very wide and very red pass band (somewhat like $R+I$), as seen in Fig.~\ref{fig:bandpass}, left.  The result is that the $R$ magnitude on a late-type M star will be set to run slower and with fewer modes than needed.
Therefore, a color correction based on spectral type can be used for these cases.

We have created a set of synthetic color corrections, and compared them to the WFS's own estimates.
We use a self-consistent set of $VRI$ magnitudes, from the USNO UCAC4 catalog\cite{2013AJ....145...44Z}, transformed from APASS $gri$ (AB).
The color corrections are shown in Fig.~\ref{fig:bandpass}, right.
Using these color corrections gives improved MagAO performance due to the AO configurations being optimized for the true rate of photons/second received by the WFS.

\subsection{VisAO Field Stop}

\begin{wrapfigure}{r}{0.58\linewidth}
	\vspace{-38pt}
	\begin{center}
		\begin{tabular}{l}
			\includegraphics[width=\linewidth]{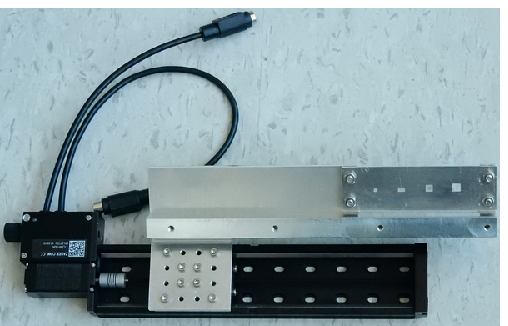}
		\end{tabular}
	\end{center}
	\vspace{-12pt}
	\caption[fieldstop]
	{ \label{fig:fieldstop}
		Stage holding new field stops enabling VisAO observations on specific targets within extended sources.
	}
	\vspace{-18pt}	
\end{wrapfigure} 

 As part of the MagAO-2K project\cite{2016males}, we have recently installed a selectable field stop mechanism for VisAO.  A linear stage at the f/16 focus prior to VisAO and the WFS moves field stops of different dimensions into the beam, facilitating better flat fielding and observations on various sources such as binary stars and protostars in H$\alpha$-glowing clouds.  The square field stops are 4$\arcs$ (intended for observations of $\alpha$ Cen\cite{2014SPIE.9148E..20M}), 8$\arcs$, and 10$\arcs$ across.  A rectangular field stop, 4$\arcs \times$8$\arcs$, prevents field-of-view overlap between the two channels in the spectral differential imaging (SDI) mode.  Figure~\ref{fig:fieldstop} shows the new stage with the field stops in the upper right of the stage.

\clearpage

\subsection{Eyepiece Observing}

\begin{figure}[htb]
	\begin{center}
		\begin{tabular}{c}
			\includegraphics[width=0.485\linewidth]{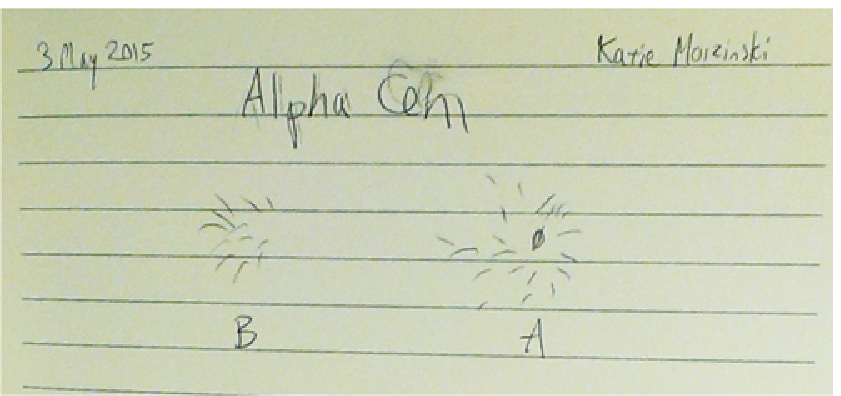}
			\includegraphics[width=0.505\linewidth]{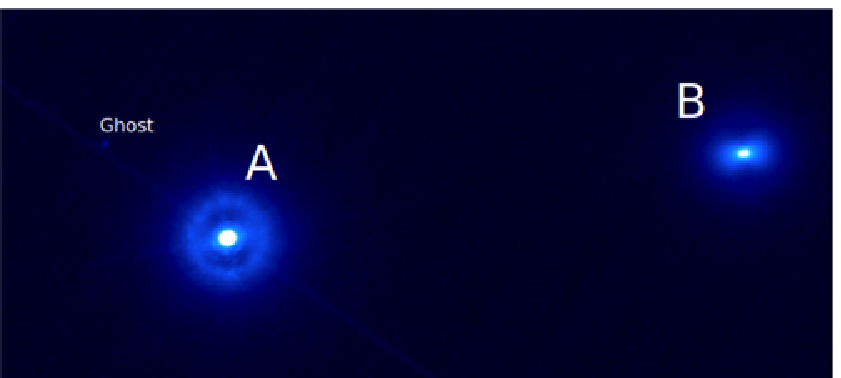}
		\end{tabular}
	\end{center}
	\caption[Alpha Cen]
	{ \label{fig:alphacen}
		$\alpha$ Cen A and B through the eyepiece \textit{(Left)} and with VisAO \textit{(Right)}.
	}
\end{figure}

On 3rd May 2015 we put an eyepiece at the usual location of the Clio dichroic, and closed the loop on $\alpha$ Centaurus A.  Figure~\ref{fig:alphacen} shows the eyepiece observation compared to an earlier VisAO observation.  The 0$\farcs$022 core and the control radius can be seen in the eyepiece observation of $\alpha$ Cen A, as well as some anisoplanatism and atmospheric dispersion are seen at the $\sim$4$\arcs$ companion $\alpha$ Cen B (there was no ADC in front of the eyepiece).  This work was reported as the NASA Astronomy Picture of the Day at \url{http://apod.nasa.gov/apod/ap150507.html}.

\section{SCIENCE WITH MagAO}

Here we give a brief survey of the 22 refereed publications that used MagAO thus far.
Four more refereed papers are currently under review or in press,
as well as 2 more submitted papers,
for approximately $\sim$25 refereed MagAO papers at the time of this writing.

\subsection{Instrumentation}

The atmospheric dispersion compensator is a unique triplet design that improves the sharpness of the visible-light AO-corrected images across the bandwidth of the VisAO filters.  This work is described in ``Design, Implementation, and On-Sky Performance of an Advanced Apochromatic Triplet Atmospheric Dispersion Corrector for the Magellan Adaptive Optics System and VisAO Camera'' by Kopon \textit{et al.} 2013 \cite{2013PASP..125..966K}.

\subsubsection{Non-Refereed MagAO Instrumentation Papers in this Proceedings}

Non-refereed papers to look for in this proceedings about vibrations on MagAO are:
``Vibrations in MagAO: resonance sources identification and first approaches for modeling and control'' by Garc{\'e}s \textit{et al.}\cite{2016garces}
and 
``Vibrations in MagAO: frequency-based analysis of on-sky data, resonance sources identification, and future challenges in vibrations mitigation'' by Z{\'u}{\~n}iga Fern{\'a}ndez \textit{et al.}\cite{2016zuniga}.

Our collaborators in Leiden had two presentations in this conference about the vector apodizing phase plate (vAPP) coronagraph we have trialed in MagAO/Clio:
``The broad-band vector Apodizing Phase Plate (vAPP) coronagraph: NIR liquid crystal performance and future developments'' by Otten
and
``Coronagraphy with two PSFs'' by Snik.
The work is more fully described in Otten \textit{et al.}, submitted\cite{otten2016}.

A paper in this proceedings regarding cloud computing using MagAO/VisAO exoplanet data is
``High-contrast imaging in the cloud with klipReduce and Findr'' by Baltzell \textit{et al.}\cite{2016baltzell}.
Additionally, MagAO was used to help simulate the GMT phasing optics in this proceedings,
``On-sky demonstration of the GMT dispersed fringe phasing sensor prototype on the Magellan Telescope'' by Kopon \textit{et al.}\cite{2016kopon}.

Finally, papers in this proceedings about the instrument include a paper describing our future MagAO upgrades and new instruments:
``The path to visible extreme adaptive optics with MagAO'' by Males \textit{et al.}\cite{2016males}
and
a paper reviewing VisAO and other visible-light AO systems: ``A Review of Astronomical Science with Visible Light Adaptive Optics'' by Close\cite{2016close}.

\subsection{Exoplanets and Brown Dwarfs}

\subsubsection{$\beta$ Pic b}

\begin{figure}[htb]
	\begin{center}
		\begin{tabular}{c}
			\includegraphics[width=0.39\linewidth]{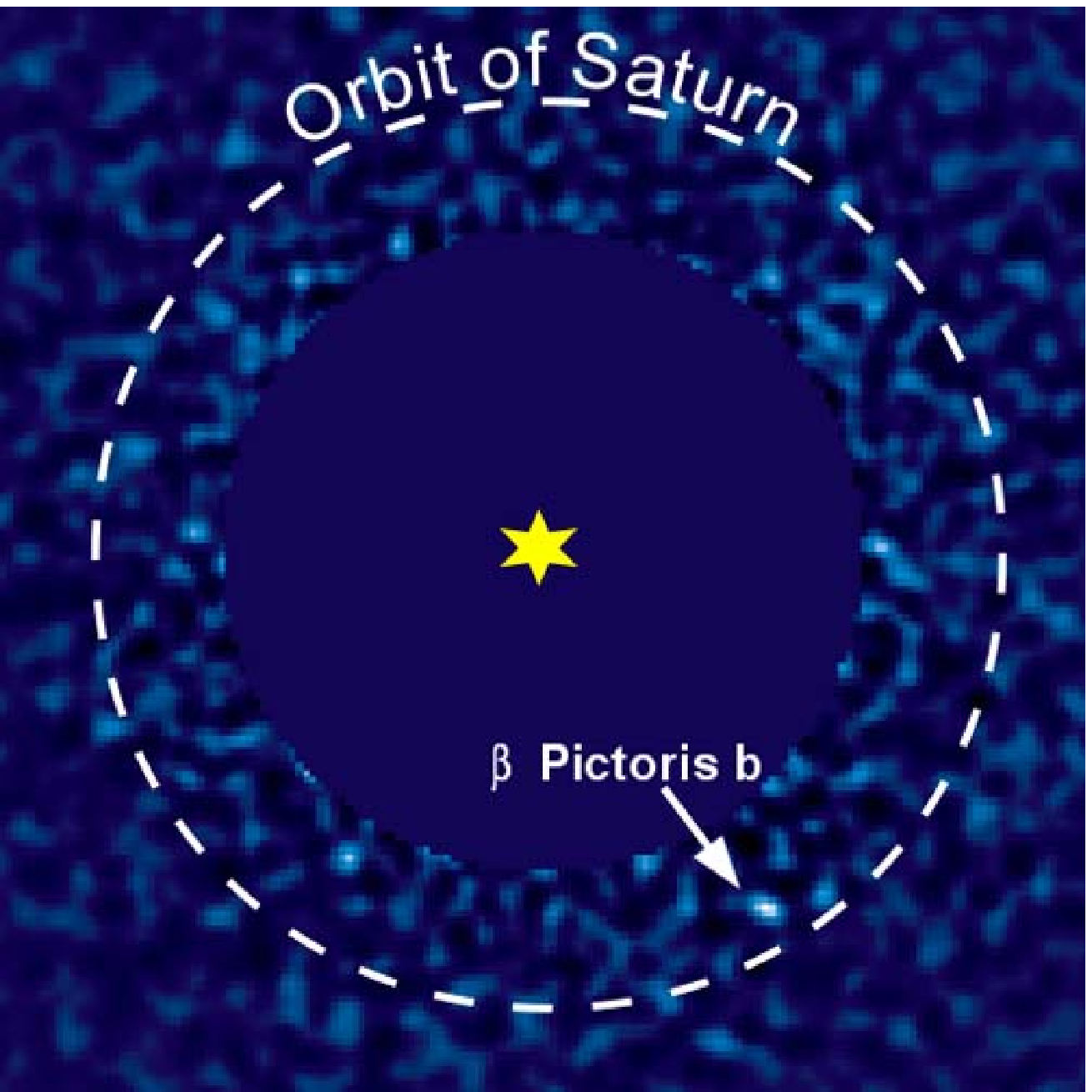}
			\includegraphics[height=0.59\linewidth,angle=90,trim=0.6cm 1cm 0.9cm 0.7cm,clip=true]{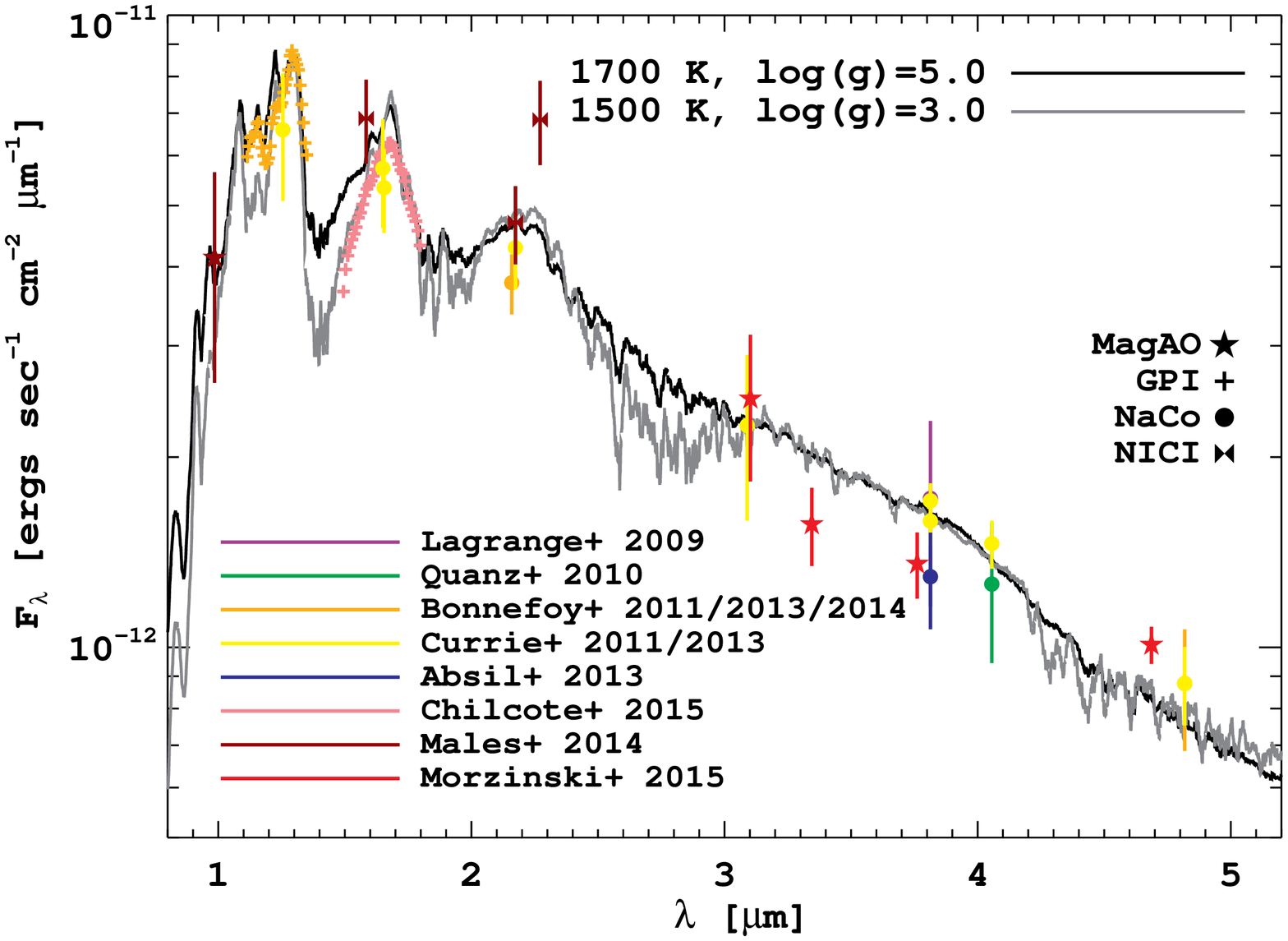}
		\end{tabular}
	\end{center}
	\vspace{-10pt}
	\caption[$\beta$ Pic b]
	{ \label{fig:betapic}
		$\beta$ Pic b results from first-light of MagAO.
		\textit{Left:} The first CCD image of an exoplanet from the ground.  Further analysis and results are given in Males \textit{et al.} 2014\cite{2014ApJ...786...32M}.
		\textit{Right:} Shown here are two best-fit atmospheric models, one with a higher temperature but lower gravity (black line), and one with a lower temperature but higher gravity (grey line).  Thus, we find that fitting atmospheric models to the spectral energy distribution is degenerate in temperature and gravity, but gives the same luminosity.  Therefore, in Morzinski \textit{et al.} 2015\cite{2015ApJ...815..108M} we elected to determine empirical rather than modeled quantities.  We measured the bolometric luminosity, empirically for the first time, finding the planet to be $\sim$20\% brighter than brown dwarfs of a similar spectral type.
	}
\end{figure}

First-light observations of known extrasolar planet $\beta$ Pictoris b were conducted in December 2012.  $\beta$ Pic b is a massive ``super-Jupiter'' (approximately 10 times the mass of Jupiter) on a Saturn-scale orbit around a nearby young star, $\beta$ Pic A.  It had never before been imaged in the visible, and the detection with VisAO (Fig.~\ref{fig:betapic}, left) inferred a lower temperature for the exoplanet than brown dwarfs at the same spectral type.  Details are described in ``Magellan Adaptive Optics First-light Observations of the Exoplanet {$\beta$} Pic b. I. Direct Imaging in the Far-red Optical with MagAO+VisAO and in the Near-IR with NICI'' by Males \textit{et al.} 2014\cite{2014ApJ...786...32M}.

We acquired thermal-infrared images simultaneously with Clio.  We combined these data with literature photometry to create the 0.9--5 $\mu$m spectral energy distribution (SED).  Attempting to fit atmosphere models to the full SED results in degeneracies between temperature and radius (Fig.~\ref{fig:betapic}, right).  However, an integration of the SED gives the bolometric luminosity, measured empirically for the first time, and also infers that the exoplanet is $\sim$20\% brighter than brown dwarfs of the same spectral type.  This work is described in more detail in ``Magellan Adaptive Optics First-light Observations of the Exoplanet {$\beta$} Pic b. II. 3-5 {$\mu$}m Direct Imaging with MagAO+Clio, and the Empirical Bolometric Luminosity of a Self-luminous Giant Planet'' by Morzinski \textit{et al.} 2015\cite{2015ApJ...815..108M}.

Both of these results together (cooler temperature yet brighter luminosity than brown dwarfs of similar spectral types) provide empirical evidence for a large radius (a.k.a.\ low surface gravity) in this young exoplanet, as predicted theoretically for young forming objects.

MagAO observations were also used to refine our measurement of the orbital parameters of $\beta$ Pic b.  The orbit is of great interest because there is a chance the exoplanet may transit (pass in front of) its star in 2017.  The MagAO astrometry is combined with Gemini/NICI data and analyzed in
``The Gemini NICI Planet-Finding Campaign: The Orbit of the Young Exoplanet $\beta$ Pictoris b''
by Nielsen \textit{et al.} 2014 \cite{2014ApJ...794..158N}.

\subsubsection{HD 106906 b}
MagAO's first new exoplanet discovery was a super-Jupiter at a very wide orbit around HD 106906.  The primary star is in fact a close binary which hosts a debris disk, making the system very interesting for understanding planet formation in diverse environments.  The discovery is decribed in
``HD 106906 b: A Planetary-mass Companion Outside a Massive Debris Disk'' by Bailey \textit{et al.} 2014\cite{2014ApJ...780L...4B}.

The system is further analyzed in
``Magellan AO System $z^\prime$, $Y_s$, and $L^\prime$ Observations of the Very Wide 650 AU HD 106906 Planetary System''
by Wu \textit{et al.} 2016 \cite{2016ApJ...823...24W}, where the first $z'$ photometry is measured and the debris disk is discovered to be particularly asymmetric at $L'$.
Moreover, Wu \textit{et al.} 2016 show that, unlike brown dwarfs CT Cha B or 1RXS 1609 B, HD 106906 b is a planetary mass object as opposed to an obscured brown dwarf.  

\subsubsection{Lk Ca 15 b}
\begin{figure}[htb]
	\begin{center}
		\begin{tabular}{c}
			\includegraphics[width=0.99\linewidth]{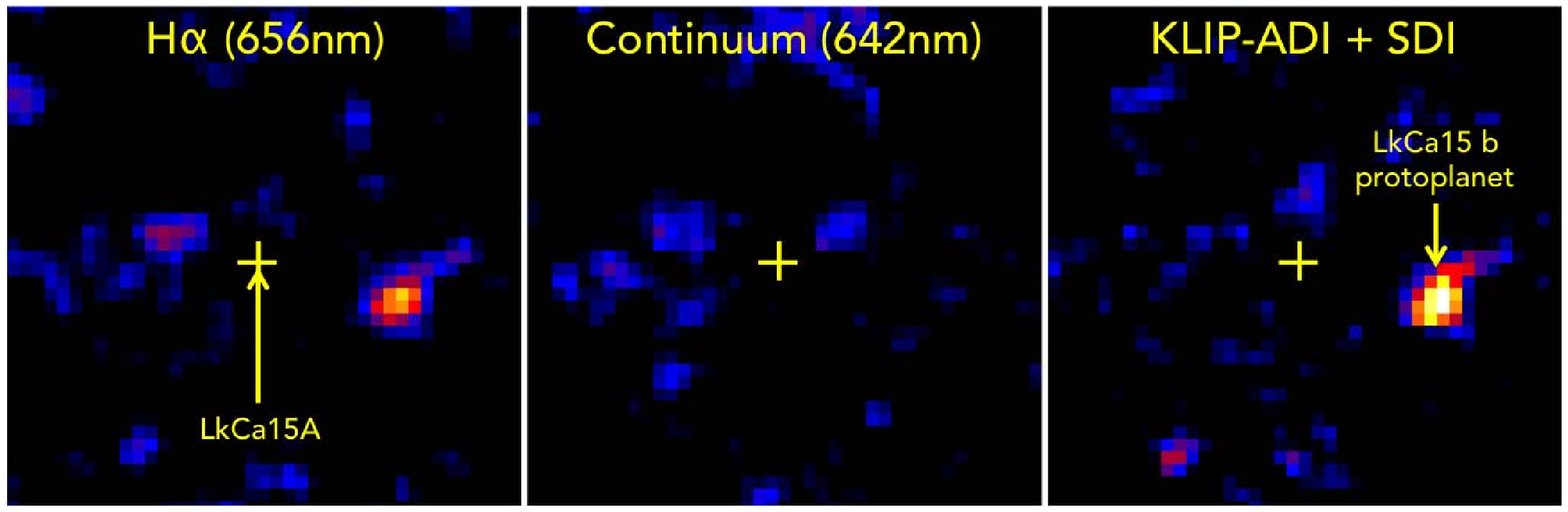}
		\end{tabular}
	\end{center}
	\caption[Lk Ca 15]
	{ \label{fig:lkca15}
		Spectral differential imaging (SDI) observations of Lk Ca 15 reveal that its companion b is accreting hydrogen, evidence that the protoplanet is currently forming.
	}
\end{figure}

MagAO/VisAO observations of the star Lk Ca 15 using spectral differential imaging (SDI) show that its closest companion is bright at H$\alpha$ (656 nm) but faint at the continuum (642 nm).  This is direct evidence for hydrogen accretion onto the object Lk Ca 15 b, indicating that it is indeed a low-mass protoplanet in the process of formation, as hinted by previous infrared observations.  The VisAO observations are shown in Fig.~\ref{fig:lkca15}.
Infrared non-redundant masking (NRM) observations at the LBT were combined with the MagAO data and reported in
``Accreting protoplanets in the LkCa 15 transition disk''
by Sallum, Follette \textit{et al.} 2015 \cite{2015Natur.527..342S}.
Further NRM results are discussed in the non-refereed paper in this proceedings:
``Imaging protoplanets: observing transition disks using non-redundant masking'' by Sallum \textit{et al.} \cite{2016sallum}.

\subsubsection{Other Brown Dwarf and Exoplanet Companions}
MagAO/Clio observations of the planetary-mass companion to the brown dwarf 2MASS 1207 in the methane band were reported in
``Directly Imaged L-T Transition Exoplanets in the Mid-infrared'' by Skemer \textit{et al.} 2014 \cite{2014ApJ...792...17S}, which included off-axis AO correction on a faint guide star.

MagAO/VisAO observations of the brown dwarf companion CT Cha B, which led to a fuller picture of the SED of the object, are reported in
``New Extinction and Mass Estimates from Optical Photometry of the Very Low Mass Brown Dwarf Companion CT Chamaeleontis B with the Magellan AO System''
by Wu \textit{et al.} 2015 \cite{2015ApJ...801....4W}.

MagAO/VisAO observations of the substellar companion 1RXS 1609 B, with a similarly improved picture of the SED, are reported in
``New Extinction and Mass Estimates of the Low-mass Companion 1RXS 1609 B with the Magellan AO System: Evidence of an Inclined Dust Disk''
by Wu \textit{et al.} 2015 \cite{2015ApJ...807L..13W}.

A method of using stellar companions as PSF stars for searching for faint companions using MagAO/Clio is described in
``Direct Exoplanet Detection with Binary Differential Imaging''
by Rodigas \textit{et al.} 2015 \cite{2015ApJ...811..157R}.

A star with a planet detected by radial velocity (r.v.) is observed to have an M dwarf companion responsible for a long-term r.v.\ trend, as reported in
``MagAO Imaging of Long-period Objects (MILO). I. A Benchmark M Dwarf Companion Exciting a Massive Planet around the Sun-like Star HD 7449''
by Rodigas \textit{et al.} 2016 \cite{2016ApJ...818..106R}.

A nearby brown dwarf triple system that could serve as a benchmark system, imaged with MagAO/Clio, is reported in
``Adaptive Optics imaging of VHS 1256-1257: A Low Mass Companion to a Brown Dwarf Binary System''
by Stone \textit{et al.} 2016 \cite{2016ApJ...818L..12S}.

\subsection{Stars and Disks}
\subsubsection{Orion}

At first light we pointed the new MagAO/VisAO to the Orion Trapezium cluster and resolved the 0$\farcs$033 binary $\theta^1$ Ori C, for the first time with filled-aperture imaging.  The images and astrometric analysis of binary star orbits in the cluster are given in
``Diffraction-limited Visible Light Images of Orion Trapezium Cluster with the Magellan Adaptive Secondary Adaptive Optics System (MagAO)'' by Close \textit{et al.} 2013\cite{2013ApJ...774...94C}.

Also in the Orion nebula, one of the photo-evaporating disks believed to be protoplanetary in nature, called ``proplyds,'' was imaged with MagAO/VisAO.  An analysis of structure due to ionization and an estimate of the central stars' masses are enabled by the observations, reported in
``High Resolution H$\alpha$ Images of the Binary Low-mass Proplyd LV 1 with the Magellan AO System'' by Wu \textit{et al.} 2013\cite{2013ApJ...774...45W}.

Some disks in the Orion nebula can be seen in silhouette as they shadow the bright nebular emission behind them.  Imaging one such disk with MagAO/VisAO revealed its structure and composition, as described in
``The First Circumstellar Disk Imaged in Silhouette at Visible Wavelengths with Adaptive Optics: MagAO Imaging of Orion 218-354'' by Follette \textit{et al.} 2013 \cite{2013ApJ...775L..13F}.

\subsubsection{Transitional Disks}

A companion at 0$\farcs$086 in a gap in the transitional disk around HD 142527 was imaged in H$\alpha$ and the continuum by MagAO/VisAO.  This result indicated accretion onto a low-mass binary stellar companion, and confirmed a tentative NRM discovery, as reported in
``Discovery of H$\alpha$ Emission from the Close Companion inside the Gap of Transitional Disk HD 142527'' by Close \textit{et al.} 2014 \cite{2014ApJ...781L..30C}.
MagAO/VisAO astrometry of the companion was also presented in ``Polarized Light Imaging of the HD 142527 Transition Disk with the Gemini Planet Imager: Dust around the Close-in Companion''\cite{2014ApJ...791L..37R}.

A point-like source seen in the HD 169142 transitional disk at $L'$ but not at H$\alpha$ with MagAO is announced and discussed in
``An Enigmatic Point-like Feature within the HD 169142 Transitional Disk''
by Biller \textit{et al.} 2014 \cite{2014ApJ...792L..22B}.

The candidate companion inside the T Cha transitional disk is imaged with MagAO/Clio and combined with data from VLT, and the system is analyzed in
``New Spatially Resolved Observations of the T Cha Transition Disk and Constraints on the Previously Claimed Substellar Companion''
by Sallum \textit{et al.} 2015 \cite{2015ApJ...801...85S}.

\subsubsection{Other Stellar and Disk Observations}

The debris disk around HR 4796A is imaged with MagAO/VisAO at $i'$, $z'$, and $Y_s$, and with MagAO/Clio at $K_s$, [3.1], [3.3], and $L'$.  Its structure and detailed constraints on its composition are described in
``On the Morphology and Chemical Composition of the HR 4796A Debris Disk''
by Rodigas \textit{et al.} 2015 \cite{2015ApJ...798...96R}.

The Wolf-Rayet star NaSt1 is imaged with MagAO/Clio and combined with \textit{HST} and \textit{Chandra} observations to deduce an extended structure around an evolved massive binary, as reported in
``Multiwavelength observations of NaSt1 (WR 122): equatorial mass loss and X-rays from an interacting Wolf-Rayet binary''
by Mauerhan \textit{et al.} 2015 \cite{2015MNRAS.450.2551M}.

\subsection{Extragalactic}

Supernovae shed light on massive star evolution.  They manifest as a bright new star seen somewhere in the sky, and one crucial observation upon discovery of a new supernova is to image the area at high spatial resolution.  Comparing ``before'' and ``after'' pictures allows for identification of which star in an historical image is the progenitor of the new supernova.
On 2016 Feb.\ 8, Type IIb supernova ``2016adj'' was discovered in the active galaxy Centaurus A.
MagAO had just been installed on the telescope for the start of the 2016A run, with the VisAO science camera mounted but not yet the Clio infrared camera.  We were able to use VisAO alone by mounting a beamsplitter at the usual location of Clio's dichroic entrance window.  In this way, we imaged the location of the supernova, and were able to tentatively identify the progenitor, as reported in
``Possible Identification of the Progenitor of SN 2016adj in NGC 5128 (Centaurus A)'' by Dyk \textit{et al.} 2016 \cite{2016ATel.8693....1D}.

\section{GLYCOL CLEAN-UP} \label{sec:cleanup}
\begin{wrapfigure}{r}{0.6\linewidth}
	\vspace{-18pt}
	\begin{center}
		\begin{tabular}{c}
			\includegraphics[width=\linewidth]{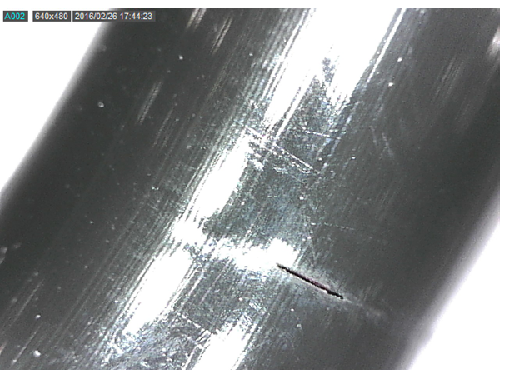}
		\end{tabular}
	\end{center}
	\vspace{-12pt}
	\caption[Hole]
	{ \label{fig:hole}
		A rupture in an internal cooling line (shown here) led to a damaging glycol leak during the 2016A run.
		The outer diameter of the hose is 8mm.
		The crack was along the outer curve of a bend, and pointed almost straight down when mounted and with the telescope at zenith.
	}
	\vspace{-18pt}
\end{wrapfigure} 

On 2016 Feb.\ 19 between 19:00-20:00 local time an internal leak of glycol coolant occurred inside the MagAO ASM while operational at the telescope.
The culprit was revealed to be a rupture in a hose that had been installed $\sim$6 years previously (see Fig.~\ref{fig:hole}).
The hose was replaced and a cleanup operation commenced.
This included cleaning the glycol contamination from the back of the thin shell, re-coating selected armatures with conductive silver for the capacitive position sensors, cleaning and repairing electronic components, and testing the ASM with and without the thin shell.

\subsection{Ultrasonic Bath}

\begin{figure}
	\begin{center}
		\begin{tabular}{c}
			\includegraphics[width=0.49\linewidth]{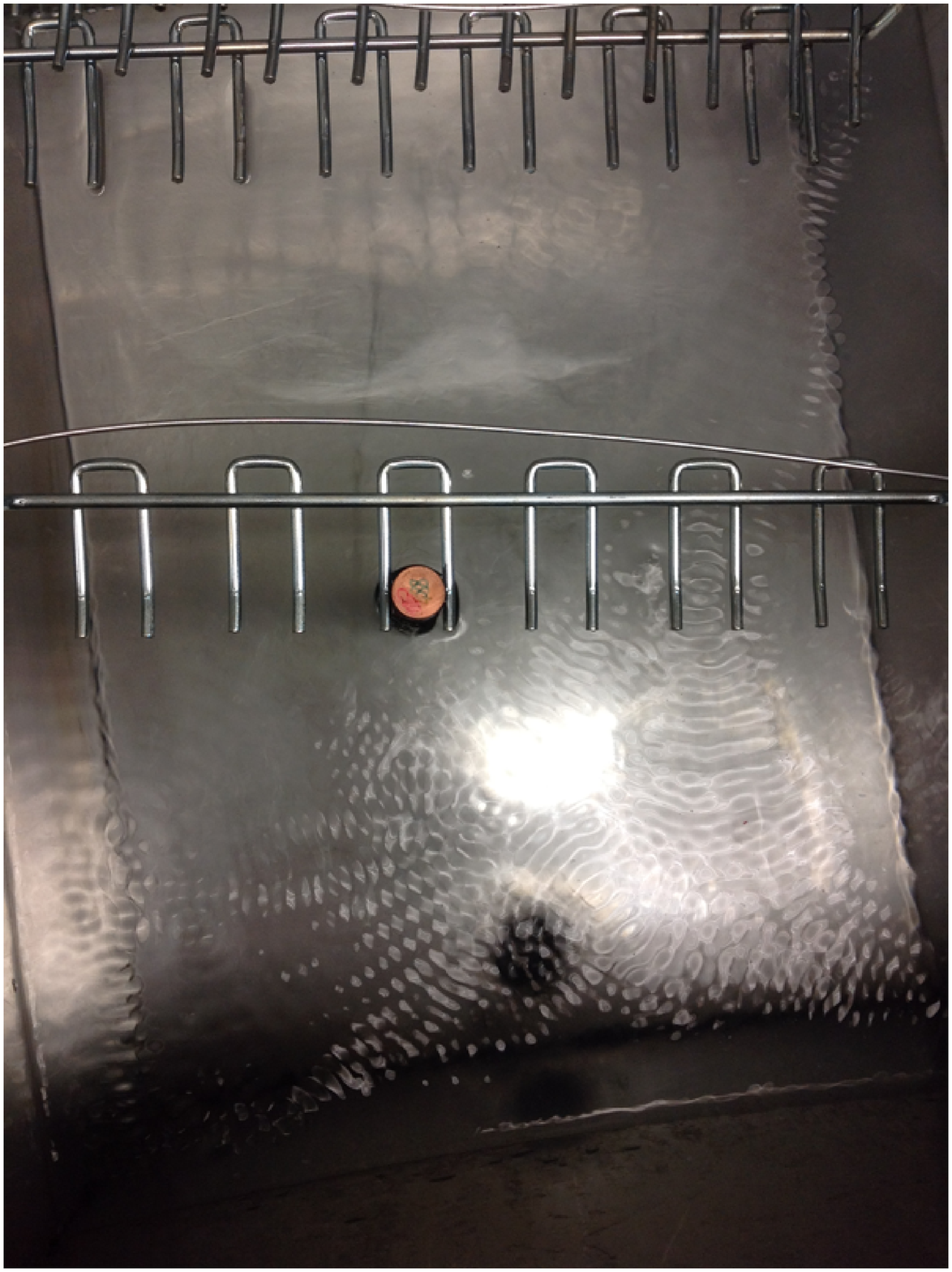}
			\includegraphics[width=0.49\linewidth]{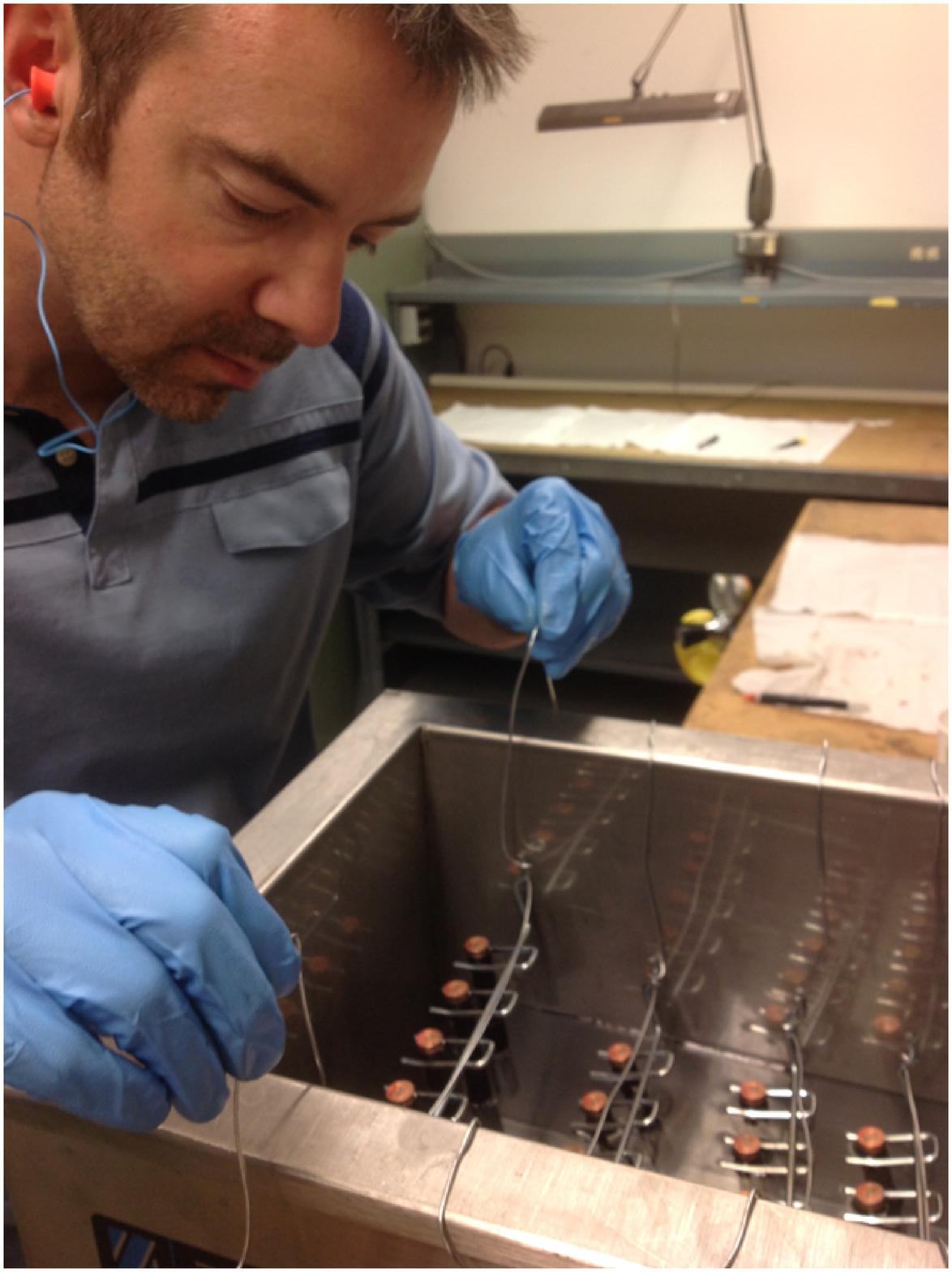}
		\end{tabular}
	\end{center}
	\caption[Ultrasonic bath]
	{ \label{fig:ultrasonicbath}
		Ultrasonic bath and the process to clean the actuators.
		\textit{Left:} A single actuator was tested first.  Waves created by the ultrasonic generator can be clearly seen.
		\textit{Right:} Batches of actuators were cleaned after the single test.
		Safety equipment included ear plugs due to the loud noise of the ultrasonic generator, and gloves to keep the actuators clean.
	}
\end{figure}

As part of this effort, we removed 41 actuators from the ASM and returned them to Arizona for cleaning.
The actuators were contaminated with glycol residue.
We used an ultrasonic bath to clean the actuators.
A tabletop Zenith ultrasonic generator/tank was used to provide the ultrasonic waveform to scrub the actuators (the waveform can be seen in Fig.~\ref{fig:ultrasonicbath}, left).
We did not use the heating feature, but operated it at room temperature.
An 80 kHz OMEGA-HF generator creates ultrasonic waves that clean small-scale features via ``ultrasonic cavitation,'' a process in which microscopic bubbles form and collapse that results in a scrubbing-action to remove contaminants from small features\cite{zenithmanual}.
A solution of 2\% soap in DI water was prepared in the ultrasonic tank.
The tank was powered on for about 30 minutes with just the soap solution inside.
This allows for de-gassing of the bubbles in the soap and water.
Then the actuators were suspended in the tank on wire supports (to keep the copper magnets from getting wet) and the ultrasonic waveform was generated for a further 30 minutes.

The procedure was as follows:
We mixed the solution, filled the tank, and de-gassed the bubbles.
We brushed the electronic components on each actuator with soap solution and a gentle bristle brush.  We suspended each actuator on a wire holder (see Fig.~\ref{fig:ultrasonicbath}) and placed the actuators into the bath with all but the actuator coil submerged.
After running the bath for about 30 minutes, we removed the actuators, loosened the screw holding the circuit card to the metal cylinder, blew the actuator dry with dry nitrogen, and laid the actuators out to fully dry.
When all the actuators were cleaned and dry, we shipped them to our partners at Microgate s.r.l., Bolzano, Italy to test and replace any broken actuators.
The result was that 31 of the contaminated actuators were salvaged, while 10 were not working and had to be replaced with spares.

\subsection{Actuator Photos}
\begin{figure}
	\begin{center}
		\begin{tabular}{c}
			\includegraphics[height=3.8in]{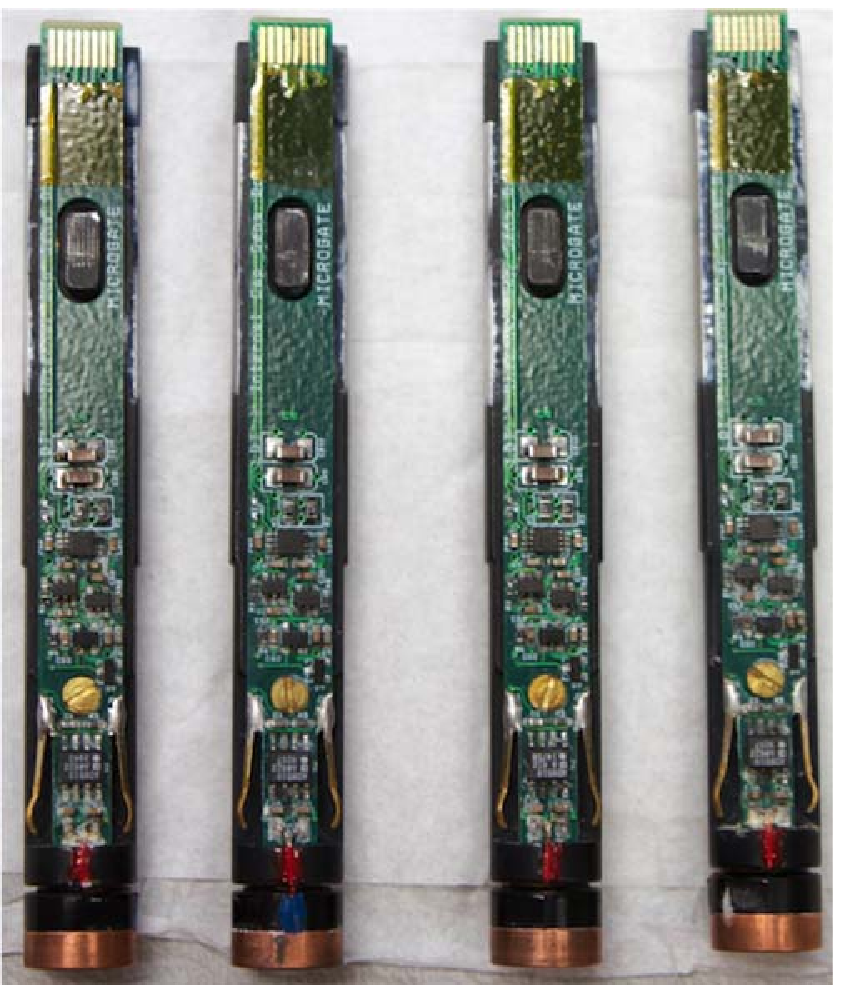}
			\includegraphics[height=3.8in]{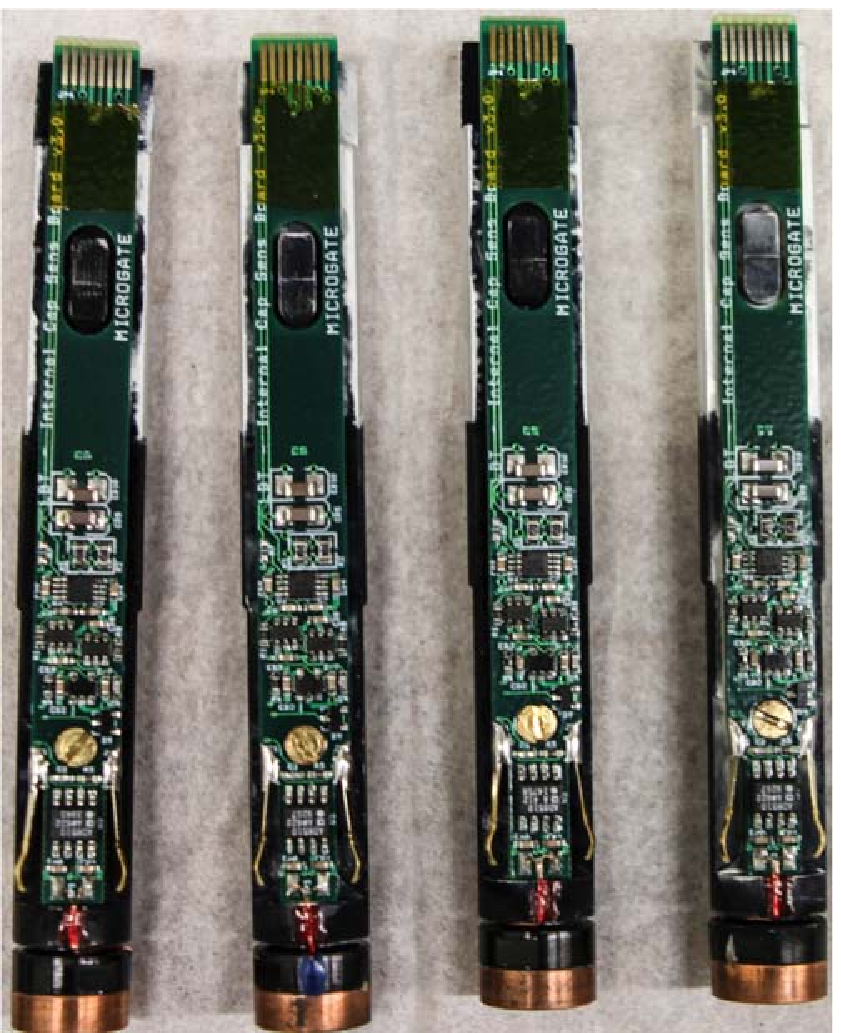}
		\end{tabular}
	\end{center}
	\caption[Before and After Actuators]
	{ \label{fig:actuators1}
		Actuators 175, 138, 67, and 229,
		before \textit{(left)} and after \textit{(right)} cleaning.
		Residue was successfully removed from all 41 actuators.
		However, only 31 of the actuators passed electrical tests after the cleaning, so 10 had to be replaced.
	}
\end{figure}

We photographed the contaminated actuators, before and after cleaning.
An example of these photos is shown in Fig.~\ref{fig:actuators1}.
The glycol residue was successfully removed by the ultrasonic bath.

\subsection{First Refurbishment Run}
In May--June 2016, a portion of the MagAO team returned to LCO to refurbish the ASM.
The 31 cleaned actuators and 10 new actuators were replaced.
41 armatures on the reference body were re-silvered, while 6 were not able to be completed.
Our somewhat novel approach of \textit{in-situ} silver ``spray'' plating of a 100-nm thick layer resulted in a new electrically-conductive coating of only $\sim$2 $\Omega$ of resistance from the inner hole onto one plate of the capacitive position sensor.
A movie of our newly-developed process can be seen at \url{https://www.youtube.com/watch?v=zeBMgkH7sj4}.

All the tight bend cooling lines (like the one that failed) were completely replaced in the ASM with new hoses by ADS of Italy.
The electronics and glycol systems were tested with the thin shell installed and found to be fully functional by Microgate.
In September 2016 a second refurbishment run will be carried out to fully integrate and test the ASM, in preparation for going back on-sky in Nov.\ 2016.

\acknowledgments     
MagAO was built with support from the NSF MRI, TSIP and ATI programs.

This material is based upon work supported by the National Science Foundation under Grant No.\ 1506818.

KMM's and LMC's work is supported by the NASA Exoplanets Research Program (XRP) by cooperative agreement NNX16AD44G.

JRM's work is supported by the NASA Exoplanet Science Institute Sagan Fellowship Program.  This work was performed in part under contract with the Jet Propulsion Laboratory and is funded by NASA through the Sagan Fellowship Program under Prime Contract No. NAS7-03001.  JPL is managed for the National Aeronautics Space Administration (NASA) by the California Institute of Technology.

Any opinions, findings, and conclusions or recommendations expressed in this publication are those of the authors and do not necessarily reflect the views of the National Aeronautics Space Administration (NASA), the National Science Foundation (NSF), the California Institute of Technology, or the University of Arizona.

We would like to thank the Las Campanas Observatory and Carnegie Observatories staff for support of the MagAO project in general and assistance with the glycol leak in particular.
We would like to thank Roberto Biasi and Mario Andrighettoni of Microgate and Enzo Anaclerio and  Daniele Gallieni of ADS for assistance with the ASM refurbishment.
We would like to thank Bryan Keener Smith of Optical Sciences, University of Arizona for coating expertise and work to coat the armatures on the reference body with silver.

Thanks to Dillon Hanrahan for conducting the NMR analysis, Tom Solsten for explaining the NMR analysis, Mary Kay Amistadi for conducting the mass spectrometer analysis, the Polt Lab at the University of Arizona's Chemistry \& Biochemistry Department for testing the pH of the glycol sample, and the University of Arizona Lab of Emerging Contaminants for analyzing the glycol sample for contaminants.
Thanks to Guido Brusa of LBTO for assistance with shipping actuators to Microgate in Italy.

Finally, we would like to thank Yuri Beletsky for documenting the eyepiece observing.

This research was made possible through the use of the AAVSO Photometric All-Sky Survey (APASS), funded by the Robert Martin Ayers Sciences Fund.
This research has made use of NASA's Astrophysics Data System.
This research has made use of the SIMBAD database, operated at CDS, Strasbourg, France\cite{2000A&AS..143....9W}.
This paper includes data gathered with the 6.5 meter Magellan Telescopes located at Las Campanas Observatory, Chile.

\appendix    
\section{GLYCOL CHEMISTRY} \label{sec:glycol}

After the Feb.\ 2016 leak,
a sample of the glycol solution was taken for chemical analysis in Arizona.
Here we report the results of these glycol tests, carried out in March--April 2016.

The chemical name is ethylene glycol and the chemical formula is HOCH$_2$CH$_2$OH.

\subsection{Concentration}

\begin{figure}[htb]
	\begin{center}
		\begin{tabular}{c}
			\includegraphics[width=\linewidth]{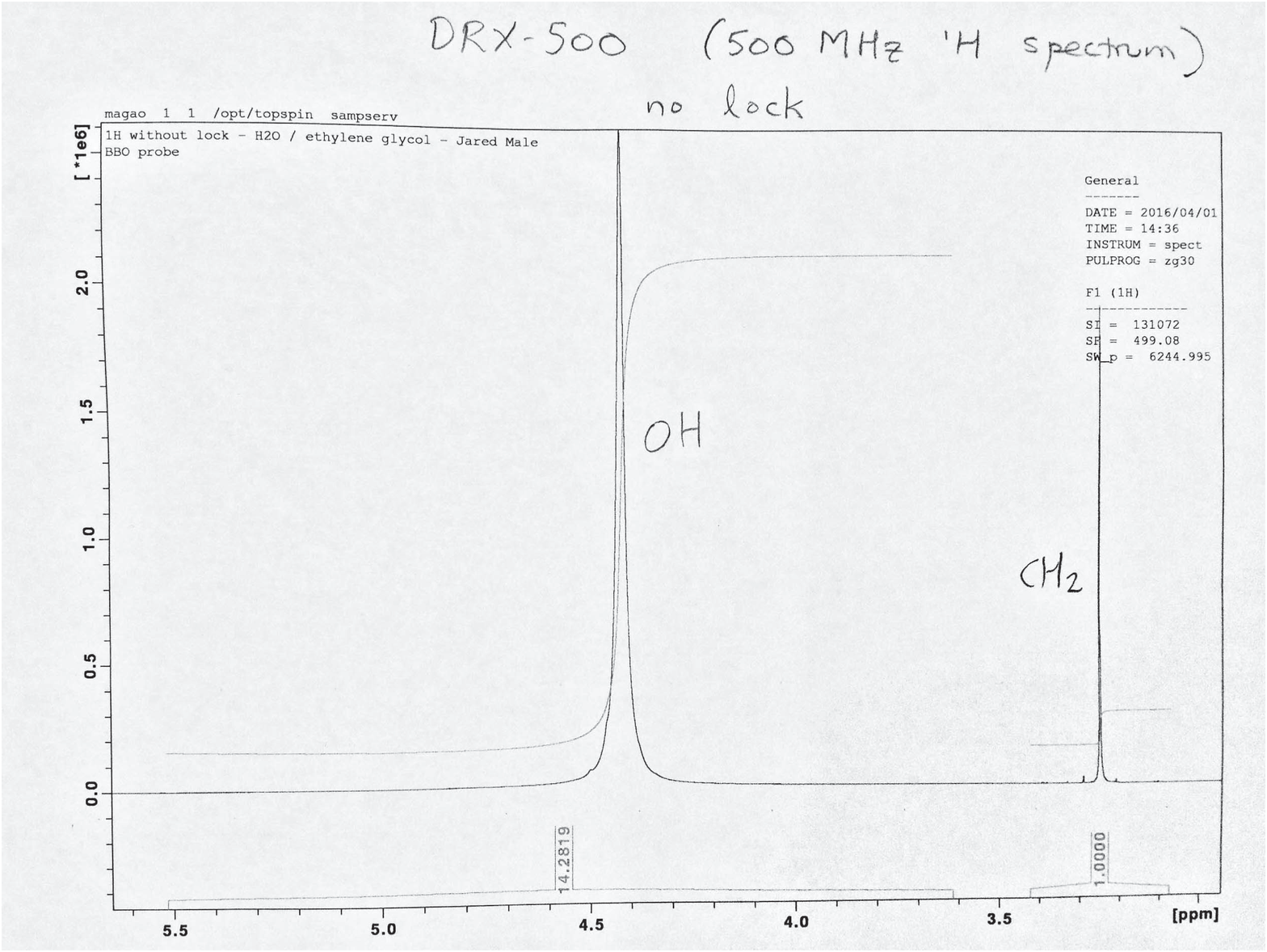}
		\end{tabular}
	\end{center}
	\caption[Glycol Chemistry]
	{ \label{fig:glycolchem}
		NMR analysis of the peaks of OH and CH$_2$ in the sample glycol-water solution.
		The ratio of the OH peak to the CH$_2$ peak is 14.28.
	}
\end{figure}

The glycol concentration in water was determined by nuclear magnetic resonance (NMR) analysis at the University of Arizona's Chemistry Department.  The NMR analysis gives a proton weight ratio, finding the ratio of the OH peak to the CH$_2$ peak to be 14.28 (see Fig.~\ref{fig:glycolchem}).
From this result, we calculate the volume mixing ratio via two methods, below.

\subsubsection{Molar calculation for pure ethylene glycol+water mixture}
Here we calculate the volume mixing ratio from the molar ratios measured via NMR.
In this calculation, we assume that the mixture only consists of pure water and pure ethylene glycol.

Let $a$ be the concentration of water, [H$_2$O].
Let $b$ be the concentration of glycol, [HOCH$_2$CH$_2$OH].
Given the chemical formulae for water and glycol,
\begin{align*}
OH =& 2a + 2b \\
CH_2 =& 4b \\
\frac{OH}{CH_2} =& \frac{2a + 2b}{4b} = 14.28 \\
2a + 2b =& 14.28 \times 4 b
\end{align*}
the molar ratio of water to glycol is:
\begin{equation}
[a/b] = \frac{(14.28 \times 4) - 2}{2} = 27.56 .
\end{equation}

The density of ethlyene glycol is 1.1132 g/L, and the density of water is 1 g/L.
Let $V_a$ and $V_b$ be the volume of water and glycol, respectively.
Let $\rho_a$ and $\rho_b$ be the density of water and glycol, respectively.
Let $M_a$ and $M_b$ be the molar mass of water (18.01528 g/mol) and glycol (62.07 g/mol), respectively.
Then the ratio by volume is
\begin{align*}
V_a / V_b =& \frac{ M_a / \rho_a } {M_b / \rho_b } \times [a/b] \\
V_a =& V_b \times 8.9 \\
\frac{V_b}{V_a + V_b} =& 0.1010
\end{align*}

Thus we find that the concentration of glycol if the solution contains only glycol and water is 10\% by volume.

\subsubsection{Using DOWTHERM SR-1 specifications}
Here we calculate the volume mixing ratio using the specifications for DOWTHERM SR-1 glycol\cite{dowtherm_sr-21}.
SR-1 is not pure ethylene glycol but contains additional agents including corrosion inhibitors.

From NMR we get a molar ratio of water to glycol of
\begin{align*}
27.56 \mathrm{\: moles \: water} :& \: 1 \mathrm{\: mole \: glycol} .
\end{align*}
The weight of water in the mixture found via NMR is
\begin{align*}
27.56 \mathrm{\: moles} \times 18.02 \mathrm{\: g/mole} =& \: 496.63 \mathrm{\: g} .
\end{align*}
The corresponding weight of glycol is
\begin{align*}
1 \mathrm{\: mole} \times 62.07 \mathrm{\: g/mole} =& \: 62.07 \mathrm{\: g} .
\end{align*}

SR-1 is specified as 95.5\% ethylene glycol and 4.5\% performance additives by weight\cite{dowtherm_sr-21}.
Let $W_g$ be the weight of glycol and $W_w$ be the weight of water found via the NMR molar ratio.
Therefore, the weight ratio given the NMR molar ratio is
\begin{align*}
W_g / (W_g/0.955 + W_w) =& 62.07/(62.07/0.955 + 496.63) \\
=& 0.1105
\end{align*}
or 11.05\% glycol by weight in the mixture.

Given this weight ratio, and again using the specifications for SR-1\cite{dowtherm_sr-21}, we find by linear interpolation that our volume ratio is 9.9\% glycol, and 10.4\% SR-1.  This value is in agreement with the 10\% value found via the first method, above.

The freezing point for this mixing ratio of SR-1 is $-3.7 ^\circ$C.

\subsection{Acidity/Alkalinity}
A member of the Polt Lab at the University of Arizona's Chemistry \& Biochemistry Department analyzed the glycol sample.
The pH was reported as 8.35 $\pm$ 0.5.
This value is as expected from the corrosion inhibitors in DOWTHERM SR-1.

\subsection{Contamination}
The University of Arizona Lab of Emerging Contaminants analyzed the glycol sample via mass spectrometry.  Contaminants found included P, K, some Na and Si, with a little Cu, and some Zn.
The full findings are reported in Tab.~\ref{tab:contamination}.

\begin{table}[!h]
	\caption{Glycol contaminants investigated via mass spectroscopy.} 
	\label{tab:contamination}
	\begin{center}
		\begin{tabular}{ll|ll|ll}
Analyte & Concentration ($\mu$g/g) & Analyte & Concentration ($\mu$g/g) & Analyte & Concentration ($\mu$g/g) \\
\hline
\hline
Li	&	not detected	&	Ga	&	not detected	&	Pr	&	not detected	\\
Be	&	not detected	&	Ge	&	not detected	&	Nd	&	not detected	\\
B	&	0.17	&	As	&	not detected	&	Sm	&	not detected	\\
Na	&	53.55	&	Se	&	not detected	&	Eu	&	not detected	\\
Mg	&	0.49	&	Br	&	0.05	&	Gd	&	not detected	\\
Al	&	not detected	&	Rb	&	0.07	&	Tb	&	not detected	\\
Si	&	9.96	&	Sr	&	0.01	&	Dy	&	not detected	\\
P	&	315.57	&	Y	&	not detected	&	Ho	&	not detected	\\
K	&	1003.19	&	Zr	&	not detected	&	Er	&	not detected	\\
Ca	&	not detected	&	Nb	&	not detected	&	Tm	&	not detected	\\
Sc	&	not detected	&	Mo	&	not detected	&	Yb	&	not detected	\\
Ti	&	not detected	&	Pd	&	not detected	&	Hf	&	not detected	\\
V	&	0.01	&	Cd	&	not detected	&	Ta	&	not detected	\\
Cr	&	0.01	&	Sn	&	0.12	&	W	&	not detected	\\
Mn	&	0.02	&	Sb	&	not detected	&	Hg	&	not detected	\\
Fe	&	0.46	&	I	&	not detected	&	Tl	&	not detected	\\
Co	&	not detected	&	Cs	&	not detected	&	Pb	&	0.50	\\
Ni	&	0.05	&	Ba	&	not detected	&	Bi	&	0.01	\\
Cu	&	4.66	&	La	&	not detected	&	Th	&	not detected	\\
Zn	&	1.90	&	Ce	&	not detected	&	U	&	not detected	\\
		\end{tabular}
	\end{center}
\end{table}


\bibliography{spie9909-1}   
\bibliographystyle{spiebib}   

\end{document}